\documentstyle[11pt,aaspp4]{article}
\tighten
\eqsecnum

\begin{document}

\font\twelvei = cmmi10 scaled\magstep1 
       \font\teni = cmmi10 \font\seveni = cmmi7
\font\mbf = cmmib10 scaled\magstep1
       \font\mbfs = cmmib10 \font\mbfss = cmmib10 scaled 833
\font\msybf = cmbsy10 scaled\magstep1
       \font\msybfs = cmbsy10 \font\msybfss = cmbsy10 scaled 833
\textfont1 = \twelvei
       \scriptfont1 = \twelvei \scriptscriptfont1 = \teni
       \def\mit{\fam1 }
\textfont9 = \mbf
       \scriptfont9 = \mbfs \scriptscriptfont9 = \mbfss
       \def\bmit{\fam9 }
\textfont10 = \msybf
       \scriptfont10 = \msybfs \scriptscriptfont10 = \msybfss
       \def\bmsy{\fam10 }

\def\etal{{\it et al.~}}
\def\eg{{\it e.g.,~}}
\def\ie{{\it i.e.,~}}
\def\lsim{\raise0.3ex\hbox{$<$}\kern-0.75em{\lower0.65ex\hbox{$\sim$}}} 
\def\gsim{\raise0.3ex\hbox{$>$}\kern-0.75em{\lower0.65ex\hbox{$\sim$}}} 

\title{The MHD Kelvin-Helmholtz Instability II:\\
The Roles of Weak and Oblique Fields in Planar Flows
\altaffilmark{10}}

\author{T. W. Jones \altaffilmark{1,2,6},
        Joseph B. Gaalaas\altaffilmark{1,2,7},
        Dongsu Ryu \altaffilmark{3,4,8},
and     Adam Frank\altaffilmark{1,5,9}}

\altaffiltext{1}{Department of Astronomy, University of Minnesota,
    Minneapolis, MN 55455}
\altaffiltext{2}{Minnesota Supercomputer Institute, University of Minnesota,
    Minneapolis, MN 55455}
\altaffiltext{3}{Department of Astronomy \& Space Science, Chungnam National
    University, Daejeon 305-764, Korea}
\altaffiltext{4}{Department of Astronomy, University of Washington,
    Box 351580, Seattle, WA 98195-1580}
\altaffiltext{5}{Department of Physics and Astronomy, University of
    Rochester, Rochester NY 14627-0171}
\altaffiltext{6}{e-mail: twj@astro.spa.umn.edu}
\altaffiltext{7}{e-mail: gaalaas@msi.umn.edu}
\altaffiltext{8}{e-mail: ryu@sirius.chungnam.ac.kr}
\altaffiltext{9}{e-mail: afrank@alethea.pas.rochester.edu}
\altaffiltext{10}{To appear in the Astrophysical Journal, June 10, 1997}

\begin{abstract}

We have carried out high resolution MHD simulations of the
nonlinear evolution of Kelvin-Helmholtz
unstable flows in $2\frac{1}{2}$ dimensions. The modeled flows and
fields were
initially uniform except for a thin shear layer with a hyperbolic tangent
velocity profile and a small, normal mode perturbation. These simulations
extend work by \cite{fran96}  and \cite{maletal96}.  
They consider periodic sections of flows
containing magnetic fields parallel to the shear layer, but projecting over a full range of angles
with respect to the flow vectors.
They are intended as preparation for fully $3D$ calculations and to
address two specific questions raised in earlier work: 
1) What role, if any, does the orientation of the field play in
nonlinear evolution of the MHD Kelvin-Helmholtz instability in $2\frac{1}{2}$D.
2) Given that the field is too weak to stabilize against a linear
perturbation of the flow,
how does the nonlinear evolution of the instability depend on strength of the field.
The magnetic field component in the third direction contributes only
through minor pressure contributions, so the flows are essentially $2$D.
In \cite{fran96} we found that fields too weak to stabilize a linear
perturbation may still be able to alter fundamentally the
flow so that it evolves from the classical ``Cat's Eye''
vortex expected in gasdynamics into a marginally stable, broad laminar
shear layer. In that process the magnetic field plays the role of a
catalyst, briefly storing energy and then returning it to the 
plasma during reconnection events that lead to dynamical alignment between
magnetic field and flow vectors.
In our new work we identify another transformation in the
flow evolution for fields  below a critical strength. That we
found to be $\sim 10$\%
of the critical field needed for linear stabilization in the cases we studied.
In this ``very weak field'' regime, the role of the magnetic field
is to enhance the rate of energy dissipation within and around the
Cat's Eye vortex, not to disrupt it. 
The presence of even a very weak field can add substantially to the
rate at which flow kinetic energy is dissipated.

In all of the cases we studied magnetic field amplification by
stretching in the vortex is limited by tearing mode, ``fast'' reconnection
events that isolate and then destroy magnetic flux islands within the vortex and
relax the fields outside the vortex. If the magnetic tension
developed prior to reconnection is comparable to Reynolds
stresses in the flow, that flow is reorganized during reconnection.
Otherwise, the primary influence on the plasma is generation of
entropy. The effective expulsion of flux from the vortex is very
similar to that shown by \cite{weiss66} for passive fields in idealized
vortices with large magnetic Reynolds numbers. We demonstrated that
this expulsion cannot be interpreted as a direct consequence of steady,
resistive diffusion, but must be seen as a consequence of unsteady
fast reconnection.

\end{abstract}


\clearpage

\section{Introduction}

Weak magnetic fields
threading conducting fluid media can play vital dynamical roles, even when
traditional criteria, such as relative magnetic and gas pressures, suggest
the fields are entirely negligible. Perhaps the best example of this is
the destabilizing influence of a vanishingly small magnetic field crossing
a Keplerian accretion disk (\cite{balhal91}), where the mere presence of the
field seems fundamentally to alter the local flow properties. 
Other examples abound, however, that
could be particularly important in astrophysics.
Among them, we would include
weak fields penetrating turbulent or otherwise strongly
unstable flows, where such fields significantly alter evolution
and transport properties,
(\eg \cite{biswel89}; \cite{catvain91}; \cite{noret92}; \cite{junetal95}). 
Sheared motion
is a critical common element in many of these flows, and the consequent stretching of
a weak, but large-scale field can lead to a locally enhanced role for
the field. The flows will also frequently lead to current sheets and
associated magnetic field
topologies unstable to reconnection, and that is central to
the nonlinear evolution of the systems (\eg \cite{bis93};~\cite{park94}).
Through these processes the fields can also have more global consequences.
Study of the nonlinear evolution of the classical Kelvin-Helmholtz (KH) instability
could be particularly useful as a well-defined example of strongly
sheared flows. Further, since KH unstable boundary layers are
probably common, the behavior of the instability is important for its
own sake. 

Although the KH instability is fairly well studied in ordinary
hydrodynamics (\eg \cite{corsher84}), comparable study has been much slower
in magnetohydrodynamics (MHD). That is
because the magnetic field substantially complicates the physics itself
and also because computational methods and resources needed for such
studies are only recently up to the task. The linear analysis of the
MHD KH instability is relatively straightforward and was long ago
carried out for a number of simple flow and field configurations
(\eg \cite{chan61}; \cite{miupr82}). 
Generally, and especially if the velocity change
is not supersonic, the ordinary fluid shear layer is unstable to
perturbations with wave vectors in the plane of the shear layer and
with wavelengths greater than the thickness of the layer (\eg \cite{miura90}).
When there is a
field component projecting onto the flow field, magnetic
tension provides a stabilizing influence. 
A simple vortex sheet is stabilized against linear perturbations 
whenever the magnetic field strength is sufficient that 
$c_a > |(\hat {\bmit k} \cdot {\bmit U_o})/
(2 \hat {\bmit k} \cdot \hat{\bmit B_o})|$,
where $\bmit U_o$ is the velocity difference between
the two layers, $c_a$ is the Alfv\'en speed, ${\bmit k}$ is the
perturbation wave vector, and $\hat{\bmit B_o}$ is the direction of the 
magnetic field (\cite{chan61}). 

\cite{fran96} (Paper I) and \cite{maletal96} (MBR)
recently presented complementary nonlinear analyses of the MHD KH instability
in mildly compressible flows based on two-dimensional numerical simulations 
carried out with new (and different) Riemann-solver-based MHD codes. 
While not the first numerical studies of
the MHD KH instability, they represented big improvements over previous
calculations in both numerical resolution and extent to which flow
evolution was followed towards asymptotic states (readers are
referred to Paper I for additional, earlier citations). 
Considering perturbed ``2D'' flows that were uniform except for a
thin, smooth velocity transition layer, those two papers emphasized the 
qualitatively different behaviors in the nonlinear evolution of unstable
flows depending on how close the field strength is to its critical strength
for stabilization. For fields only slightly below the critical value,
enhancements in the tension of the field through linear growth can
stabilize the flow before it develops distinctly nonlinear characteristics. 
For weaker fields, however, the
initial evolution of the instability is very similar to that for the
ordinary KH instability. That results in the formation of eddies, and hence
to substantial stretching of the magnetic field lines as well as
reconnection. Paper I emphasized the remarkable fact that in a case
with a field  2.5 times weaker than critical, reconnection can lead to
self-organization in the flow and fairly rapid relaxation to a
quasi-steady laminar and marginally stable flow. MBR presented summaries
of simulations extending to somewhat weaker fields showing evidence for
similar behaviors.

Neither Paper I nor MBR, however, explored the problem in
sufficient depth to establish the conditions necessary for the
previously mentioned self-organization. In addition, it is very important in
this situation to understand how the magnetic fields behave when they
are ``very'' weak (a concept whose definition needs clarification, in fact).
A closely related matter is what differences, if any, exist between
the behavior of a truly weak field and a stronger field whose projection onto
the flow vectors is weak. Alternatively stated, are there differences
between the nonlinear ``2D'' MHD KH instability and the ``$2\frac{1}{2}$D'' MHD
KH instability? Answers to those basic questions are the objective of this
paper. 
We find: 1) for the cases we have considered with an initially
uniform field that the magnetic field transverse to the plane is
unimportant and 2) there is a transition from the role of the
magnetic field as a catalyst to flow self-organization to a role as
an added source of energy dissipation that should vanish directly as the
initial magnetic field strength projected onto the plane vanishes.
Ultimately we must understand the full ``3D'' problem, in which
the perturbation wave vector also lies outside the flow direction.
On the other hand it has been difficult to carry out 3D MHD simulations
with sufficient numerical resolution to be confident of the results in
complex flows such as these.
In addition, it will be useful to compare fully 3D behaviors with
2D flows.
We hope that the current work is a significant, constructive step towards
a full understanding of this problem.

The paper plan is as follows. In \S 2 we will summarize the problem
set-up and relevant results from Paper I. \S 3 contains a discussion of
new results, while \S 4 provides a brief summary and conclusion. We also
include an appendix presenting an analytical model for diffusive
flux expulsion from a steady vortex, in order to contrast that physics with
what we observe in the eddies that form in our simulations.

\section{Background}

In order to focus on specific, important physical issues we have chosen
to explore an idealization of the MHD KH problem, reserving for
the future the more general problem. We present in this section
only a bare outline
of our method and some key results from previous work. A full discussion
of the computational setup along with several tests of such issues as
adequate numerical resolution and geometry of the
computational box can be found in Paper I.

\subsection{Problem Definition}

The geometry of the computations is shown in Figure 1. The only difference
from Paper I is that in those earlier computations
we assumed an aligned field,
$\theta = \cos^{-1}{|\hat {\bmit B_o}\cdot{\bmit \hat U_o}}| = 0$,
whereas we now relax that 
constraint to include magnetic fields oblique or orthogonal to the flow plane. 
We assume the flow to be periodic in $x$ and that the $y$
boundaries are reflecting
(\ie neither flow nor field lines cross the $y$ boundaries). This was a
configuration used initially by \cite{miura84}, and we followed it
in Paper I to enable a direct comparison with his results. The
influences of those boundary choices are discussed fully in Paper I.
In brief, the periodic boundary limits coalescence of structures to scales
equal to the box dimension, $L$. That is more significant than the
existence of the reflecting boundaries, which seem to have only minor
influence on dynamics in the (narrow) central regions where flow organization and
dissipation is largely determined.

The initial background flow has uniform density, $\rho = 1$,
gas pressure, $p = 0.6$, and an adiabatic index, $\gamma = 5/3$,
so that the sound speed, $c_s = \sqrt{{{\gamma p}/{\rho}}}=
1.0$.  The magnetic field, 
${\bmit B_o} = B_o ( \hat {\bmit x}\cos{\theta}  +
\hat {\bmit z}\sin{\theta} )$, 
is also uniform.
In Paper I we considered cases with $B_o = 0.4,~0.2$ (and $\theta = 0$),
so that $M_A = 2.5,~5$, since those were studied by \cite{miura84}.
We now add to these a number of new cases as outlined in Table 1. 
To facilitate
comparisons we identify the simulations from Paper I as
Cases 1 and 2 in Table 1,
with the new simulations following.
The velocity in the background state is antisymmetric about $y = L/2$
according to the relation
\begin{equation}
{\bmit u_o} =u_o(y)\hat {\bmit x}  = ~-~{U_o
\over 2} \tanh\left({y - {L/2} \over {a}}\right)\hat {\bmit x},
\label{vprof}
\end{equation}
with $U_o = 1$.
This describes a smoothly
varying flow within a shear layer of full width $2a$. 
For all our simulations presented here $a = L/25$, chosen to make the
interactions with the reflecting boundaries small. The square computational
box has $L = 2.51$.
Flow is to the left in the top half-plane and to the right below that.
To this state we add a perturbation, $\delta (\rho, p, \bmit u, \bmit B)$,
defined to be a normal mode found from the linearized MHD equations
appropriate for the chosen background, periodic 
in $x$ and evanescent in $y$, with period equal to the
length of the computational box, $L$. 
This was done exactly as in Paper I. 
We note that when $\theta \ne 0$ there are perturbations in all
three vector components of the magnetic and velocity fields.
Under the conditions we used the
computational frame is comoving with the KH waves.
All flow and field quantities are either symmetric or antisymmetric
around two points, which happen with our choice of phases in 
$\delta (\rho, p, \bmit u, \bmit B)$ to be at $y = L/2$, $x = L/4,~3L/4$. 
Because the velocity field is antisymmetric around these points they are
the places where strong vortices tend to form.

The equations we solve numerically are those of ideal compressible MHD; namely,
\begin{equation}
{\partial\rho\over\partial t} + {\bmsy\nabla}\cdot
\left(\rho {\bmit u}\right) = 0, 
\label{masscon}
\end{equation}
\begin{equation}
{\partial{\bmit u}\over\partial t} + {\bmit u}\cdot{\bmsy\nabla}
{\bmit u} +{1\over\rho}{\bmsy \nabla}p - {1\over\rho}
\left({\bmsy\nabla}\times{\bmit B}\right)\times{\bmit B} = 0, 
\label{forceeq}
\end{equation}
\begin{equation}
{\partial p\over\partial t} + {\bmit u}\cdot{\bmsy\nabla}p
+ \gamma p{\bmsy\nabla}\cdot{\bmit u} = 0, 
\label{energy}
\end{equation}
\begin{equation}
{\partial{\bmit B}\over\partial t} - {\bmsy\nabla}\times
\left({\bmit u}\times{\bmit B}\right) = 0, 
\label{induct}
\end{equation}
along with the constraint ${\bmsy\nabla}\cdot{\bmit B}=0$ imposed to
account for the absence of magnetic monopoles (e.g., \cite{priest84}).
The isentropic gas equation of state is $p\propto \rho^{\gamma}$.
Standard symbols are used for common quantities.  Here, we have
chosen rationalized units for the magnetic field so that the magnetic
pressure $p_b = B^2/2$ and the Alfv\'en speed is simply $c_a =
B/\sqrt{\rho}$.

These equations were solved using a multidimensional MHD
code based on the explicit, finite difference ``Total Variation
Diminishing'' or ``TVD'' scheme.  The method is an MHD extension of
the second-order finite-difference, upwinded,
conservative gasdynamics scheme of \cite{harten83},
as described by \cite{ryuj95}.  The multidimensional version
of the code, along with a description of various one and two-dimensional
flow tests is contained in \cite{ryujf95}.
The code contains an {\it fft}-based
routine that maintains the ${\bmsy\nabla}\cdot{\bmit B}=0$ condition
at each time step within machine accuracy.
That step does {\it not} compromise the other conservation relations.

Numerical solution of equations (\ref{masscon}) - (\ref{induct}) on
a discrete grid leads, through truncation errors, to diffusion of energy,
and momentum, as well as
to entropy generation. Of course, such effects are also present
in nature and are important to defining the character of the flows.
The existence of effective numerical resistivity is necessary, for
example, to allow magnetic reconnection to occur in the calculations. Our 
methods exactly conserve
total energy, as well as mass, momentum and magnetic flux,
so the exchange between
kinetic, thermal and magnetic energies along with entropy production is
internally consistent.
There is fairly good evidence that conservative
monotonic
schemes, as this one is, do a good job of approximately representing
physical viscous and resistive dissipative processes that are expected
to take place on scales smaller than the grid ({\it e.g.,} \cite{porwood94}).
For the astrophysical environments being simulated the expected
dissipative scales are likely very much smaller than those that can be modeled
directly.
Recent numerical studies of reconnection suggest in the MHD limit with large
kinetic and magnetic Reynolds numbers that the local energy
dissipation rate through reconnection becomes independent of the value of the
resistivity in complex flows (\eg \cite{bis93};~\cite{bis94}).
However, when we depend on numerical dissipation, we must be cautious 
about the possible role of ``uncaptured'' dynamical structures that
could be expected on scales smaller than the grid scale, or about magnetic field
structures on such scales that could enhance the number of reconnection sites. 
In fact, we shall see that for our KH instability-induced flows involving weak
fields (where reconnection
topologies are formed on many scales) {\it total energy
dissipation is slightly ``enhanced''
when the grid is finer}. This is opposite to what we expect from
the effects of reduced numerical diffusion alone,
but consistent with an increase in the number of reconnection sites,
or ``X'' points, allowed when smaller-scale field structures can be
resolved cleanly..
Further, we find that for a fixed numerical resolution, addition of
a very weak magnetic field substantially enhances the rate of energy dissipation,
again as one expects in response to reconnection (\eg \cite{zim96}).
Thus, we see strong evidence for
unsteady, local reconnection as in high Reynolds number, MHD turbulence,
but also that our numerical solutions are not quite converged in terms
of total dissipation.

Below we will express all of our results in time units defined by the
growth time of the linear instability, $t_g = \Gamma^{-1}$, as estimated
from graphs presented by \cite{miupr82}.  
That is, we express time as $\tau = t/t_g$.
We find that especially the
initial saturation of the instability, but also the relaxation
processes are fairly uniformly expressed in these units.  
The time units, $t_g$, are listed for each case in Table 1, along with
the duration of the simulation in units of $\tau$.
The flows examined in the present paper all have $t_g \approx 1.6 - 1.7$.
For comparison the sound crossing time in the box is $t_s = L = 2.51$
(since $c_s = 1$) and the Alfv\'en wave crossing time is $t_A = M_A t_s$
(since $U_o = 1$).
The normalized turnover time for a large eddy is roughly $t_E \sim L/(U_0/2) \sim 2 t_s$.

\subsection{Magnetic Field Evolution}

Since it enters prominently into our later discussions, it is helpful
here to remind readers in a simple way of what we can expect for the
local evolution of the magnetic field in these simulations. A full analysis 
of all the subtleties is beyond our scope here, so readers
are referred to detailed discussions such as those
by \cite{moff78}, \cite{priest84}, and \cite{bis93}.
Equation (\ref{induct}) follows fields when the resistivity is exactly
zero. As already mentioned, our finite difference code will introduce
effects that mimic a finite resistivity, although it is not possible
to define an exact value for the resistivity, $\eta$. The effective
resistivity will also depend on grid resolution, decreasing roughly as
$N^{-2}$ within smooth flows. 
Despite these limitations we can make good heuristic use of
the resistive MHD extension of the induction equation (\ref{induct}).
For our purposes 
it is interesting to cast that equation in the following form:
\begin{equation}
\frac{d~\ln{(|B|/\rho)}}{dt} =\frac{1}{2} \frac{d~\ln{p_b}}{dt} - \frac{d~\ln{\rho}}{dt} =
\frac{{\bmit B}\cdot [({\bmit B}\cdot{\bmsy\nabla}){\bmit u}] -
\eta j^2 + \eta {\bmsy\nabla}\cdot({\bmit j}\times {\bmit B})}{B^2},
\label{dbdt}
\end{equation}
where $d/dt$ is the Lagrangian time derivative, the first term 
on the right represents field amplification by stretching, and
the last two terms containing the resistivity account for magnetic
``annihilation'' and ``diffusion''. 
We have also used equation \ref{masscon}.
The current density, ${\bmit j = {\bmsy\nabla}\times B}$.
The term $\eta j^2$ is a dissipative
term that balances the Joule heating in the analogous energy equation
for the gas (\eg Paper I). 
The last term is written as an expression involving the Lorentz force,
${\bmit j\times B}$, to show that it represents the transport of momentum
flux in response to resistivity; \ie the slippage of field lines.

A frozen-in field results when $\eta = 0$.
That leaves only the time derivatives and one term on the right of equation \ref{dbdt}.
If there is flow compression or expansion only perpendicular to ${\bmit B}$
then the right side of equation \ref{dbdt} vanishes and leads to
$|B|/\rho$ = constant or $p_b \propto \rho^2$. Those are the most
common statements of field compression.
However, more generally one needs to include the other ideal MHD
term on the right that accounts for ``stretching''. 
In fact, field enhancements due to compression are much more
limited than those due to stretching, especially in mildly compressive
flows, such as those we are studying.

We also can see from equation (\ref{dbdt}) that resistive influences on
magnetic energy are associated with both Joule heating and momentum
transport. In fact reconnection leads to both irreversible heating
of the local plasma and to its acceleration.
The physics of reconnection is complex and beyond the scope of this
paper. However, it may be helpful to estimate the dissipation rate 
using equation \ref{dbdt} and the Sweet-Parker description of 
reconnection (\eg \cite{bis94};~\cite{park94};~\cite{zim96}).
Assuming an incompressible flow steadily carrying oppositely
directed fields into a current sheet of thickness, $\delta$,
equation \ref{dbdt} gives us a magnetic energy
annihilation rate and associated dissipation
rate per unit volume, $Q = \eta j^2 \sim \frac{1}{2} \frac{u}{\delta} B^2$, where
$u$ is the inflow speed. In this picture plasma flows out from the reconnection region
at the Alfv\'en speed, so mass conservation leads to the relation between
the current sheet thickness, $\delta$, and its width, $l = \delta\frac{c_a}{u}$.
Then, $Q \sim \frac{1}{2}B^2\frac{c_a}{l}$.
So, the integrated dissipation rate through reconnection depends on
the field energy advected into reconnection sites and the summed 
volumes of all the reconnection sites.
The current density within the current sheet can be estimated as $j \approx B/\delta$, so that
the aspect ratio of the reconnection region is $\delta/l \sim 1/\sqrt{N_L}$,
where  $N_L = (l~c_a)/\eta$ is known as the Lundquist number of the plasma
and is obviously related to the magnetic Reynolds number.
Consequently the dissipative volume scales as $l^2/\sqrt{N_L}$.
In 2D, dissipative reconnection regions form
out of tearing mode instabilities within a current sheet when the
aspect ratio, $\delta/l$, is small (\cite{bis94}), and, hence when the magnetic
Reynolds number is large. Thus, reconnection is  not really
steady, and the number of sites and their individual volumes will
depend on the magnetic Reynolds numbers in the flow. Those are, indeed,
the behaviors we see in our simulations.

\subsection{Issues}

We already alluded in the introduction to the basic
character of nonlinear MHD KH instability
properties found from previous work. Our intent here is to explore more
fully the behaviors of weak magnetic fields in this situation. 
In preparation for that we note from past
work several key features for 2D symmetry: 

1) When there is no magnetic field or if the field is orthogonal to the
flow direction a shear layer will ``role up''. For the periodic flows considered 
here the result is a stable ``cat's eye''
vortex whose length equals the imposed
periodicity on the space and whose height $\sim 1/3$
the length. As long as the flow is subsonic or ``submagnetosonic''
no shocks are involved and the vortex decays only through viscous diffusion.
We will not deal with the supersonic or supermagnetosonic cases here
(for some work on those see, \eg \cite{pedw91}; \cite{miura90} and references
therein).

In many astrophysical applications the
kinetic Reynolds number of the flow is very large, so we wish
to consider similar cases, so that dissipative decay times are long. 
For our simulations, the empirical viscous decay time of the flow
is at least four orders of magnitude longer than the duration of our
computations (see Case 5 energy evolution curves in Figure 3).
Thus, the cat's eye represents
a ``quasi-steady relaxed state'',
in which the shear layer has spread vertically
by horizontal localization of vorticity and become stable. Of course, for
non-periodic systems there will be continued spreading due to additional
vortex mergers, while in 3D the flows will be unstable to perturbations
directed along the third direction. Those influences are beyond the
scope of our present investigation, however.

2) In the other extreme, if the magnetic tension force produced
by a perturbation
in the shear layer exceeds the ``lift'' force produced by the perturbation,
the perturbed flow is stabilized. For wavevectors
aligned with the flow that condition exists in a linear perturbation
whenever the field is strong enough that
$M_{A\parallel} = U_o/(B_o\cos{\theta}) < M_{Ac} = 2$. 
Then only viscous diffusion contributes to spreading of the original
shear layer. As before, we can neglect that influence over finite times
and describe the shear layer
as remaining in a quasi-steady relaxed state from the start.
For the present computations the critical magnetic field to stabilize
a linear perturbation is $B_c = 0.5$.

For fields slightly weaker than critical, a small, but finite amplitude
perturbation may lead to the same stabilization condition, perhaps
after a small amount of quasi-linear growth to the instability. Such was
the result obtained for the so-called ``strong-field'' case of Paper I,
where $M_A = 2.5$. That calculation is
listed as Case 1 in Table 1, here. Under
these conditions the flow is never far from laminar and there is a
modest amount of spreading in the shear layer before it also reaches
a quasi-steady relaxed state consisting of a broadened, laminar
shear layer (a point made both in Paper I and
by MBR). 
In either Case 1 or Case 2, the total magnetic energy changed
very little during the flow evolution. In Paper I we pointed out
that for the symmetry imposed the mean vector magnetic field is
time invariant; \ie $\langle {\bmit B}\rangle =~$ constant, so any
relaxed state with a relatively uniform field will automatically
contain magnetic energy close to that of the initial conditions.
This condition just reflects the conservation of magnetic flux on the
grid.
The ``relaxed'' magnetic energy is very slightly enhanced, because the
field is not quite uniform at the end. 
A small amount of kinetic energy dissipation takes place in Case 1 before the flow
becomes relaxed. That amount is mandated by 
total energy conservation, the approximate magnetic energy conservation
in this case, and the fact that on a fixed space the kinetic energy in a broad, symmetric
shear layer is less
than in a thin one. So, any evolution leading to a broadened, laminar
shear layer requires an increased thermal energy
determined by the final width of the layer, independent of how it got there. 
In our idealization the mass in the box is also exactly conserved, so
the mean density is constant. Consequently, to first order there is
no change in thermal energy by way of reversible, adiabatic processes.
Most of the increased thermal energy must result from
entropy-producing, dissipation of some kind. Putting it simply, for
these flows to relax entropy must be generated. These energy considerations 
apply to all of the calculations in our study if the relaxed state
is a laminar flow.
(It turns out that the kinetic energy of the cat's eye vortex is less
than that for the initial flow, as well. So, its formation must also
generate entropy.)

3) For initial fields too weak to prevent formation of the cat's eye vortex
through magnetic tension, the shear layer will role up as for unmagnetized
fluid flow. In Paper I, however,
we saw that when the initial field is only a few times weaker than the
critical value for linear stabilization, there follows a dramatic
transformation in the flow as the cat's eye develops. Paper I
considered a flow with $M_A = 5$. We list it in the current Table 1 as
Case 2. As the vortex roles up in such cases,
magnetic field lines are stretched
around it, thus increasing magnetic energy at the expense of kinetic
energy of the flow. The greatest magnetic pressures are produced in thin
flux tubes formed between the vortex and its twins in periodic
extensions of the space (see Figures 4a \& 6). 
Especially within the vortex and around
its perimeter, this evolution leads to magnetic reversals, beginning
after about one turn-over time for the vortex. The reversed fields
are unstable to tearing mode reconnection. So, once that happens
the magnetic field quickly reorganizes itself, releasing the
magnetic stresses and stored magnetic energy.
For Case 2 this leads, as well, to disruption of the cat's eye and
eventually to an almost steady, laminar flow, 
after formation and disruption of several weaker vortices.
That, final, quasi-steady relaxed state,
was similar to the initial conditions, except that the shear layer
was broad enough to be stable against perturbations on scales that
fit within the periodic box.
The relaxed shear layer had a linear velocity profile (see MBR for
similar points).
There are additional interesting
characteristics of the relaxed state. Remaining fluctuations
in the magnetic and velocity fields were almost exactly correlated, 
so that they could be described as linearly polarized Alfv\'en waves.
The magnetic energy returned to a level slightly above the initial
conditions, with the final excess representing a pair of apparent
``magnetic flux tubes'' bounding a hot central core of the shear layer containing
most of the entropy generated during the relaxation process. This final
condition was reached by about $\tau = 20$. The relaxed shear layer
was broader in this case than Case 1 (see also MBR), so that the
kinetic energy was also smaller. Thus, even accounting for the slight
increase in magnetic energy, we could correctly predict that 
{\it Case 2 with a weaker field was necessarily more dissipative than Case 1}. 
That extra dissipation could come about only through the effects of reconnection.
Other aspects of this case will be visited presently, since we will encounter
them again in some new cases examined here.

Thus, from the calculations reported in Paper I and in MBR it is obvious that
weak magnetic fields can play a major role in the evolution of 
MHD KH unstable flows. It is also apparent that these roles involve
the exchange of energy and momentum
from the gas to the magnetic field and then back to the gas through
Maxwell stresses and reconnection. But, it is not yet clear what are
the crucial steps in that exchange, nor how it depends on the initial
strength of the field, so that the ordinary KH behavior results, if it does,
in the limit that the magnetic field becomes vanishingly small. In
addition, although the linear MHD KH instability is not affected
by the presence of a (possibly strong) field transverse to the
flow (but still aligned to the plane of the shear), a field 
oblique to the flow can carry ``circularly'' polarized Alfv\'en waves.
Since we found that the 2D version of the problem generated linearly
polarized Alfv\'en waves, it may be important to see if the
nonlinear problem depends at all on the orientation of the 
field, or just on the strength of the field projected onto the
plane as in the linear problem.

\section{Results}

To meet the objectives at the end of the previous section
we have carried out a set of seven new simulations. They are
outlined in Table 1 as Cases 3 - 9. (Once again, Cases 1 and 2
were discussed in Paper I.) The new simulations were designed to cover a wide
range of strengths for the magnetic field in the computational plane.
They include flows
that have exactly the same total field strength as Case 2, but in
which the initial field is oblique to the computational plane; \ie
$\theta\ne 0$ (Cases 3, 5 and 6), as well as flows in which the field
is entirely in the computational plane, but is weaker than that in Case 2
(namely Cases 4, 7, 8 and 9). The plane-projected field
strength for Cases 6 and 7 is an order of magnitude weaker than for Case 2. 
Note that Cases 3 and 4 are paired to have the same initial planar
field strengths, as are Cases 6 and 7.
This enables us to compare efficiently any
distinct roles of field strength and orientation with respect to the
plane. In Case 5 the field is orthogonal to the plane, so we expect (and see)
no important role for the magnetic field. As others have noted before,
compressible influences are controlled in that case by magnetosonic waves rather
than pure sound waves, so there is a very slight modification in response
to that ($M^{-1}_{ms} = \sqrt{M^{-2} + M^{-2}_A} = 1.02$ in Case 5).
We will not concern ourselves at all with flows in which the
field is strong, by which we mean situations where magnetic tension
precludes the nonlinear development of the MHD KH instability.

Figures 2 and 3 provide a broad overview of the evolution of the new
models we computed. They illustrate the time variation of
energy components (thermal, kinetic and magnetic), as well as
the pressure minimum ratio, $\beta_{min} = (p/p_b)_{min}$, on the grid at each time. 
Figure 2 displays results for the runs that were computed with $N_x = 512$
while Figure 3 shows the cases computed with $N_x = 256$.
In addition, to provide a sense of the influence of numerical resolution for
very weak field cases
both Case 6h ($N_x = 512$) and Case 6m ($N_x = 256$) are shown together
in Figure 2. 
Resolution issues were discussed in detail for Cases 1 and 2 in Paper I,
where similar energy plots were also given. We shall add a few
additional relevant comments below.

There is one distinctive detail about Figures 2 and 3
that is important to their interpretation. It was 
apparent to us by comparing animations of the important dynamical
quantities that Cases 3 and 4 were
virtually indistinguishable from one another,
and likewise for Cases 6m and 7.
Thus the important issues are somewhat easier to see if we eliminate
$B_z$ to first order in Figures 2 and 3.
We plot a reduced energy for the
cases with $\theta \ne 0$. The magnetic and kinetic energy plotted include
only the planar components; \ie $E'_b = \int (1/2) B^2_p dxdy$,
where $B_p = \sqrt{B^2_x+B^2_y}$,
and $E'_k = \int (1/2)\rho(u^2_x + u^2_y)dxdy$.
To compensate, the total energy is also reduced
as $E'_{Tot} = E'_b + E_t + E'_k$,
where $E_t$ is the thermal energy. Thus, we ignore the almost
constant energy contributions from $B_z$ and $v_z$.
Note, however, it is the total energy, not the reduced total
energy, that is conserved.
In the definition of $\beta_{min}$,
total magnetic pressure from all three components was used, since 
it is the entire pressure that exerts a force on the plasma.

We can see that the reduced energy evolution in
Cases 3 and 4 are almost identical in Figure 2, and the same is true of
Cases 6m and 7 in Figure 3. Thus, we observe in $2\frac{1}{2}$D,
at least, that the transverse magnetic field component ($B_z$)
plays no significant role in nonlinear evolution of the instability.
The rationale for some role comes from the observation that
finite $B_z$ and $v_z$ enable circularly polarized Alfv\'en waves 
in $2\frac{1}{2}$D, while only linearly polarized Alfv\'en waves are allowed in 2D.
In other words, there are twice as many degrees of freedom in $2\frac{1}{2}$D for
Alfv\'en waves to help disperse perturbations.
However, there are two arguments to support our observation of a minimal role.
First, the group velocity vector for Alfv\'en waves along which physical
information propagates is ${\bmit v_g}={\bmit B}/\sqrt{\rho}$; \ie
aligned with the magnetic field (see, \eg \cite{landau84}).
With invariance of quantities along the $z$-direction, only the group velocity
components projected onto the $x-y$ plane are relevant, and they are 
independent of $B_z$.
The second point is that even though $B_z$ and $v_z$ are finite in $2\frac{1}{2}$D, 
the above symmetry restricts their contributions
to those of magnetic pressure (see equations [\ref{masscon}]
- [\ref{induct}]); \ie
to the total pressure gradient and the fast and slow wave speeds.
In the present case with weak fields, even that has little importance.
For example, the evolution of $\beta_{min}$ is almost the same
for Cases 3 and 4, where stretching of the field in the plane is
dominant in Case 3. For Cases 6m and 7 the values of $\beta_{min}$
are always large, because neither the planar nor the transverse
fields are very strong. But at $\tau \sim 5$ when the
stretching of the planar field is greatest, the values of $\beta_{min}$
are still similar.
So, our results indicate that for weak fields the extra degrees of freedom allowed in $2\frac{1}{2}$D
have no significant effect on the KH instability. These flows are
essentially 2D.

In one of the most striking findings from these simulations, 
we find two distinctive evolutionary behaviors for weak field
flows, depending on the field strength. 
When the planar magnetic fields are very weak, it turns out that the
evolution of the 2D MHD KH instability is qualitatively similar to the
gasdynamic version of the instability. That is, the cat's eye vortex 
continues to exist as long as we extend the simulation. The magnetic
field does, through unsteady reconnection, substantially
enhance energy dissipation of the
vortex over that from the gasdynamical case, however. 
This can be seen by comparing Case 5 ($B_p = 0$) in Figure 3 
with Case 6m or Case 7 ($B_p = 0.02$).
We will characterize 
such flows as having ``very weak fields'', and the role of the 
field as ``dissipative''. On the other hand,
if the initial fields approach the critical
field for linear stabilization within a factor of a few,
the cat's eye is disrupted and the field causes the flow to reorganize 
into a laminar form (as in Case 2). 
This we call simply the ``weak-field'' regime, and the field as 
``disruptive''.
In either situation the magnetic energy peaks
about the time of the initial reconnection instability.
For the disruptive cases
there is a fairly prompt return of the magnetic energy near to its
initial value (see Figures 2 and 3). 
In effect, the magnetic field plays the role of a catalyst, storing
energy temporarily and through it modifying the plasma flow. That
role is fairly dramatic, since reconnection leads to dynamical
alignment and self-organization in the flow.
For dissipative cases, the magnetic energy declines much more
slowly, and in the high resolution Case 6h, seems mostly to oscillate.
That behavior results from the fact that the cat's eye vortex continues
to capture magnetic flux, amplifying it and thence dissipating it.

The existence of two qualitatively distinct evolutionary patterns
is also quite apparent in the histories of the thermal and kinetic
energies displayed in Figures 2 \& 3. The kinetic energy decay in Case 6
is almost exponential, with a time constant that can be roughly
estimated as $\tau_d \sim 10^4$. By contrast Cases 3 and 4 show 
before $\tau \sim 10$ a sharp drop in $E'_k$ along with an
accompanying increase in $E'_t$,
followed by a slow, possibly exponential evolution that
is similar to Case 6. For all cases in Table 1 the magnetic
energy is always small (although it can briefly increase by factors
between 6 and 20 it never contributes more than a few percents to the
total energy), so on the face of it the magnetic energy would
not seem to be important. In fact, it can be crucial, as we have already outlined and will
discuss more fully, below.
Figure 3, which shows a wider range of projected field strengths, 
shows similar patterns with some variation in
the abruptness of the early energy transition. 
We also conclude that as $B_p \rightarrow 0$ the evolution of
the 2D MHD KH instability will smoothly approach that
for ordinary gasdynamics.

\subsection{Very Weak Fields: Dissipative}

Let us now describe in more detail the characteristics of the two
weak-field regimes and also establish the physical boundary between them.
Case 6 and Case 7 had the weakest finite planar field, $B_{po}$,
that we considered, and they demonstrate very-weak-field patterns.
Figure 4 summarizes
flow properties for Case 6h at times $\tau = 5$ and 30. Figure 4a
shows the flow vorticity component out of the plane,
$\omega_z = ({\bmit\nabla\times u})_z$,
while Figure 4b shows the magnetic field lines as they project onto the
plane. Except for minor fine structure that aligns with the planar field, the 
vorticity at $\tau = 5$ is
the same as for a flow with $B_p = 0$ of Case 5. It also is fairly
close to that for all our simulations, except Case 1 at this early time .
Primarily this image illustrates the formation
of the cat's eye and how that concentrates vorticity. Note that the
``corners'' of adjacent cat's eyes overlap, with a shear layer
between them. It is in that shear layer where magnetic fields are
most strongly enhanced, as seen for this time in Figure 4b. Extensions
of that feature around the perimeter also contain concentrations of
magnetic flux. These regions represent flows where gas is accelerated 
out of a stagnation point midway between vortex centers.
Frozen-in fields are thus pulled out or ``stretched'' and amplified.
At $\tau = 30$, the cat's eye vortex is still
largely the same, except for
some complex, low level
vorticity structures outside the main vortex. Their origins are made clearer
by examination of the magnetic field structure at this time. Those
same regions outside the cat's eye also contain isolated
magnetic flux islands
and field reversal regions.  Such magnetic features reveal an environment
where magnetic reconnection is active. The relationship is that reconnective
processes not only reorganize the magnetic field topology and release  
magnetic energy, but they also accelerate the local plasma, and that
contributes to the local vorticity. By contrast to Case 2 (or as we shall see
Cases 3 and 4, as well), however, magnetic stresses in Cases 6 or 7 produce
only small modifications to peripheral flows, and are far too weak 
to disrupt the cat's eye. 

The vortex interior shows initial signs in Figure 4b of
magnetic flux expulsion by $\tau = 5$, a well-known phenomenon 
(\eg \cite{weiss66}). 
At this early stage, just as the vortex is fully formed, there is 
still some magnetic flux that threads  
through the ``eye''. That does not seem to be the case at the later
time. In a close examination of field structures within the
vortex we can find no evidence after about $\tau = 10$ that any magnetic
flux threads the vortex. Instead, the field breaks into flux
islands within the vortex, and those are annihilated through mergers. 
This is, of course, just what high Reynolds number (both kinetic
and magnetic), nonlinear resistive 
MHD flows are expected to
do  in response to reconnection (\eg \cite{bis93};~\cite{bis94}).
The field structure in the bottom panel of Figure 4b is qualitatively
very similar to that found by Weiss (\cite{weiss66}) from a classic passive 
field simulation in a steady vortex with magnetic Reynolds number $R_m = 10^3$, but
very different from that for small magnetic Reynolds number ($R_m = 20$). 
Our own estimates of effective magnetic Reynolds numbers in the Case 6h simulation
give numbers in excess of $10^3$ (\cite{ryujf95};~\cite{fran96}), so
the comparison is very reasonable.

We emphasize, however, that {\it the field behaviors seen here, 
and presumably by Weiss, result
from localized, inherently time dependent reconnection, not simple flux diffusion}.
To demonstrate that we compute in the Appendix the equilibrium
passive magnetic fields for a simple vortex in a resistive fluid.
Figure 5 displays two examples of magnetic field structures predicted by this
steady-state resistive MHD theory. The velocity structure for this model
vortex is similar to that observed in the simulated flows for Case 6.
The core, $r < r_0/2$, has a constant vorticity, $A$, (see Figure 4a)
while there is an outer
flange in which the velocity decreases to zero at $r = r_0$.
Solutions depend only on the effective
magnetic Reynolds number, $q = 2Ar^2_0/\eta$, within the vortex. This
parameter roughly measures the ratio of the timescale for
magnetic field diffusion to the rotation period of the vortex.
When $q$
is a few ($q = 10$ in the top panel of the figure), flux diffuses quickly 
enough to almost flatten out field lines to their external pattern
except in the vortex core. At the center field lines are actually 
concentrated by rotation into a quasi-dipole pattern.
That dipole results because reconnection into a simpler topology
is not permitted.

The steady-state pattern is very different when $q$ is large, since
field is almost frozen into the vortex.
Above $q \sim 10^2$ the field forms a spiral pattern in the vortex
core whose pitch depends on $q$. Shear is greatest in the flange
region of this vortex
and flux becomes concentrated increasingly into
the vortex perimeter.  That particular behavior
does resemble Case 6 in some ways.
There are multiple field reversals in the perimeter, with that number
determined by the balance between field diffusion and addition of
new turns to the field through rotation.
The bottom panel in Figure 5 shows the steady-state solution for $q = 10^3$.
Despite some superficial similarities,
there is an obvious and 
essential difference between the truly steady-state solution
in Figure 5 and the ``quasi-steady''
field in the two time
dependent simulations (ours and Weiss').
For the steady solutions the field line that
was originally through the vortex center remains and is wrapped into
a spiral pattern within the core giving an almost constant
field strength there. 
In the time dependent cases, reconnection
changes the field structure inside the vortex.
The same statement applies to Weiss' simulation. If the
effective magnetic Reynolds number within the flow is
large enough for topologies to develop that are tearing-mode
unstable, field within the vortex ``breaks'' off from the external
field and is destroyed.
Thus, we find good agreement with Weiss in the nature of eddy flux
expulsion and see that it cannot be viewed as a steady magnetic
diffusion.

The value of $B_p$ never drops to zero inside the vortex of any
of our very weak field cases,
although it is as much as two orders of magnitude
smaller than along the vortex perimeter. Action around the
perimeter episodically injects new magnetic flux into the
vortex interior, so this process continues as long as we have
followed the flow. Examination of the magnetic energy, $E'_b$, for
Case 6 in Figure 2 shows related, episodic peaks on rough 
intervals, $\Delta \tau \sim 2.5~-~3$.
That corresponds to about half a turn over-time for the vortex, and represents
the interval on which fields along the vortex perimeter are
stretched until they become
``folded'', so as to make them subject to tearing mode reconnection.
So, whatever flux is caught in this flow becomes amplified,
increasing the magnetic energy, before reconnection rearranges
the field lines. Those lines outside the vortex are relaxed
towards the initial field configuration, often through multiple
reconnection events, while
those inside are isolated into closed islands.
Note that the total magnetic flux threading our box
is exactly conserved and exactly zero in these computations. Thus
any field line entering on the left boundary must at all times extend
continuously to the right boundary or exit again on
the left. Closed flux islands
can exist, but nothing prevents them from being destroyed, since
they contribute no net flux.

Reconnection, and most obviously flux-island destruction, is irreversible, 
so that it must be accompanied by entropy generation. That outcome is
very apparent in Figures 2 and 3. Recall that to first order total
thermal energy changes in these, closed-system simulations reflect nonadiabatic
processes. Compare first the evolution of thermal energy in Cases 6m, 7
and 5. We see that after some viscous dissipation necessary to form the cat's
eye, the Case 5 with $B_{p0} = 0$ has almost constant total thermal
energy to the end of the simulation. By contrast Cases 6m and 7
show a steady rise in $E_t$, with small amplitude oscillations associated
with major reconnection events as described above. Thus, the main
dynamical impact of the field is enhanced dissipation.
The total magnetic energy remains small in all cases, but for
the dissipative cases that reflects a near balance between the
rate at which kinetic energy is being transferred to magnetic energy
and the rate at which magnetic energy is being dissipated. For example,
from initial field energy evolution in Figure 3 we can crudely
estimate for Case 6m that magnetic energy is generated at a rate $d~E'_b/dt \sim 3\times 10^{-4}$, 
which is very close to the mean slope of the thermal energy
curve, $E_t$.

Looking next at the Case 6m,h  $E_t$ plots in Figure 2, we see that
energy dissipation is greater in the higher resolution Case 6h.
That results despite a smaller numerical resistivity for the higher
resolution simulation.
This is because in Case 6h a larger amount of flux is caught in the
vortex, so more reconnection sites develop.
We estimate that $\sim10\%$ of the total magnetic flux 
is attached to the vortex at the end of the
calculation $(\tau=50)$ in Case 6h, while only $\sim3\%$ is attached
at the end of the calculation $(\tau=30)$ in Case 6m.
As the effective magnetic Reynolds number increases there is a 
tendency for decreased resistivity to be countered by an increase
in the number of reconnection sites, as our earlier discussion of
reconnection theory would suggest.
If we compare simple estimates for rates of magnetic energy
generation for Cases 6m and 6h to the rates of thermal energy
increase, we see that they are consistent, as we found earlier
for Case 6m, alone.
The increased dissipation for the higher resolution simulation
is relatively modest, however.
Some studies of resistive MHD turbulence suggest, in
fact, that at very large Reynolds numbers total reconnection rate
in a complex flow
will be insensitive to the value of the resistivity (\cite{bis93};~\cite{bis94}).
If confirmed more generally, that could provide a practical measure 
of convergence in studies of the present kind.

\subsection{Weak Fields: Disruptive}

Looking again at thermal energy evolution in Figures 2 and 3 we can
see a clear behavioral transformation in the sequence: 
Case 5 $\to$ 6(7)$\to$ 8 $\to$ 9 $\to$ 3(4).
This grouping is arranged in order of increased $B_{p0}$. There is a
sharp rise through the sequence in the amount of dissipation associated with the initial
formation of the cat's eye vortex, as well as a more modest increased slope to the
subsequent long-term dissipation rate. The physical character of this transformation
is apparent if one compares Figure 4b with
Figure 6. The latter shows for Case 3 the magnetic pressure distribution
at three times, $\tau = 5, 10$ and $30$. 
Behaviors for this simulation are qualitatively similar to 
Case 2, as discussed in detail in Paper I and outlined in \S 2.3. Here we note that,
while the field appears wrapped around the vortex at $\tau = 5$,
it has a laminar appearance at $\tau = 30$. The velocity field
undergoes a similar transition; \ie the vortex is
completely disrupted. The magnetic field at the intermediate
time, $\tau = 10$, shows aspects of both the other times. Curiously,
however, the flow in the dominant vortex pattern there has the opposite vorticity
to the original flow.  That feature is short-lived.
We can understand this flow transition from rotational to
laminar by examining Figures 7,
8 and 9, also relating to Case 3. In Figure 7
we show the log of $\beta = p/p_b$ with magnetic field lines
overlaid at $\tau = 5$ .
The minimum $\beta = 0.55$ in the strong flux tube connecting
vortices, but the magnetic field is dynamically significant
most of the way around the perimeter of the vortex.
The maximum $\beta \sim 10^6$ in small regions where magnetic reconnection
has begun and the field strength has decreased to very small values.

To understand how the magnetic field disrupts the vortex consider
the forces involved. 
The centripetal force associated with motion around the cat's eye is
$\rho u^2_{\phi}/R \sim \frac{1}{2}\rho U^2_0 \sim \frac{1}{2}$,
since we observe that $u_{\phi} \sim \frac{1}{\sqrt{2}}$ and $R \sim 1$
for the vortex. Indeed we also confirm that when $B_{p0} = 0$, the
pressure gradient force within the vortex, ${\bmit -\nabla}p \approx 0.5$ and is directed
towards the vortex center, so that it supplies the necessary force.
On the other hand in Figure 8
we display the magnetic field lines along with
vectors representing the magnetic tension force; \ie
\begin{equation}
{\bmit T} = ({\bmit B\cdot\nabla}){\bmit B},
\label{tension}
\end{equation}
at $\tau = 7.5$ for Case 3.
It is apparent that magnetic tension forces are
concentrated where the field has been pulled into loops by the
flow, and that they are directed 
towards the center of the vortex.
The peak value of $|T| = 1.85$. 
At $\tau = 7.5$ in Case 3, however, the total pressure gradient is
small and actually of the wrong sign to effect significantly the motion 
of the plasma. The total force vector field is very similar to the
tension force field shown in Figure 8.
At the same time ``X-points''  in the field topology show that reconnection
is underway that will isolate the associated field lines. 
Subsequently, flux islands are formed along the ``axis'' of the cat's
eye, and the magnetic tension pulls
the plasma frozen into those loops towards the original center of
the vortex, disrupting its rotation.
Field line segments reconnecting outside the vortex core will tend to relax
towards the original field topology.

These observations allow us to estimate simply what minimum initial
field, $B_{p0}$ should lead to vortex disruption.
Since the early evolution of a weak field is self-similar, we can use
the behavior from Case 3 to estimate $|T| \approx 1.85~(B_{p0}/0.14)^2$
at the time of the first major reconnection in all weak-field cases.
Our earlier discussion requires $|T|\ge \frac{1}{2} U^2_0\rho = \frac{1}{2}$
for disruption, leading to the constraint $B_{p0} \ge 0.05$, or 
alternately, $B_{p0} \ge 0.1 B_c$, where $B_c$ is the critical field
for stabilization of original instability.
Indeed, as Figure 3 demonstrates, the transformation between
dissipative and disruptive evolution occurs for conditions between
those of Cases 8 and 9; \ie for initial field values between $0.04$ and $0.07$.

Figure 9 illustrates why the vortex disruption process is also highly dissipative. The top panel
shows at $\tau = 7.5$ in Case 3 the gas entropy distribution,
while the lower panel displays the electric current density, 
$|j|$ with the field lines overlaid.
From this we can see that excess entropy is concentrated into
regions where reconnection is currently active (highlighted by
$|j|$) or recently active. Ohmic heating  ($\propto j^2$) is partly responsible for
the irreversible energy exchange. The remainder should be viscous
dissipation of small-scale, disordered motions. As we discussed
in Paper I, the final laminar flow that results in this class of
flow includes a central sheet of hot gas containing
most of the excess entropy produced through the self-organization
of the flow.

\section{Summary and Conclusion}

We have carried out a series of high resolution MHD simulations of Kelvin-Helmholtz
unstable flows in $2\frac{1}{2}$ dimensions. All of these simulations 
involve magnetic fields initially too weak to stabilize the flows in the
linear regime; \ie $B_{p0} < B_c$.
Thus, since simulations are performed on a periodic space,
flows all begin formation of a single ``cat's eye'' vortex. If the
field lying in the computational plane is absent or ``very weak''
the cat's eye structure becomes a persistent, stable feature that
represents a ``quasi-steady'' equilibrium.  When there are ``very
weak'' magnetic fields in the plane they become wrapped into the
vortex and amplified by stretching. However, within a single turn of the vortex
they are subject to tearing mode instabilities leading to 
magnetic reconnection. That reconnection isolates some magnetic
flux within the vortex, which is eventually annihilated. This
is the process through which flux is effectively expelled from
a vortex. As long as the vortex persists this process will
repeat. Since reconnection is irreversible, this process is also
dissipative and leads to an increase over viscous effects
in conversion from kinetic
to thermal energy. 
We find in this regime that as the initial magnetic field within the
computational plane is
increased the dissipation rate increases in a similar manner.
Likewise, as we use a finer numerical grid, thus {\it reducing the
effective numerical resistivity and viscosity, the dissipation rate
increases}, reflecting the increased ability of our
code to capture small-scale reconnection events. This trend
is backwards from
what one would expect if simple magnetic diffusion were primarily
responsible for the reconnection. It suggests, perhaps that if
we had been able to extend these calculations to even higher
resolution the energy dissipation rate might have converged to a
value independent of the effective resistivity, just as some
studies of resistive MHD turbulence find.
The reconnection and expulsion of flux within vortices in our
simulations are similar to those in a classic study  by Weiss of vortex flux
expulsion in large-magnetic-Reynolds-number flows.

If the initial magnetic field is strong enough that within a
single turn of the vortex it is amplified around the vortex
perimeter to ``dynamical'' strength
($B^2 \sim \rho U^2$), then the reconnection described in the
previous paragraph releases stresses that are capable of
disrupting the vortex entirely. This can happen in a single event
or, if the field is only marginally strong enough ($B_{p0} \sim \frac{1}{10} B_c$), through a succession
of dynamical realignment events. In either case the net result is
a laminar, but marginally stable flow, in which the original shear layer is 
greatly broadened. 
Thus, as we discussed fully
in Paper I, such fields can have a remarkable stabilizing
influence. This is despite the fact that their total energy content 
is a minor fraction of the total, so that they are nominally too weak to
be important, according to the usual criteria.

We considered cases in which the magnetic field was entirely within
the flow plane and others in which the field was oblique to that plane,
in order to examine the role in nonlinear flows of the component
out of the plane. For the $2\frac{1}{2}$ D flows we have studied, only
the field components in the flow plane have any dynamical significance.
In fully 3 D flows, however, we expect further evolution of the 
``quasi-steady relaxed states''
of both very weak field (or dissipative) cases and weak field 
(disruptive) cases.
The cat's eye vortex of very weak field cases is subject to a
3 D instability known as the elliptical instability (\cite{pierr86};
\cite{bayly86}) unless the flow lines around the vortex follow
perfect circles.
The planar shear flow of weak field cases is stable against
linear perturbations but unstable to 3D finite-amplitude
perturbations (\cite{boh88}).
Thus, it will be important to extend the present study to the
fully 3 D regime, and we are preparing to do that.

\acknowledgments
This work by TWJ, JBG, and AF was supported in part by the NSF through
grants AST-9318959 and INT-9511654, through NASA grants NAGW-2548 and NAG5-5055 and
by the University of Minnesota Supercomputer Institute.
The work by DR was supported in part by Seoam Scholarship Foundation.
We are grateful to B.I. Jun for stimulating and helpful discussions
about these results.

\appendix

\section{Diffusive Flux Expulsion from a Steady Vortex}

In order to show that the flux expulsion from vortex shown in
Figure 4b requires localized, unsteady reconnection, not just diffusion,
here we study the steady state passive field solutions of resistive MHD.
Under 2D and $2\frac{1}{2}$D symmetries, the planar magnetic field can be
written as ${\bmit B}={\hat{\bmit z}}\times{\bmsy\nabla}\psi$ where
$\psi$ is the magnetic flux function.
Then, the induction equation in (\ref{induct}) with constant
resistivity becomes
\begin{equation}
{\partial\psi\over\partial t} + {\bmit u}\cdot{\bmsy\nabla}{\psi}
=\eta\nabla^2\psi.
\label{magfluxeq}
\end{equation}

We consider the evolution of an initially uniform magnetic field,
${\bmit B}=B_o{\hat{\bmit x}}$ or $\psi=B_o r e^{i\phi}$, within a
vortex.
We approximate that the vortex has azimuthal velocity
\begin{equation}
v_{\phi} = \cases{2Ar, &for $0<r<r_o/2$\cr
2A(r_o-r), & for $r_o/2<r<r_o$\cr
0, & for $r>r_o$\cr},
\label{vphieq}
\end{equation}
and zero radial velocity.
$A$ equals the (constant) vorticity within the vortex core.
This is roughly the velocity field of the vortex in Figure 4.
Then, the evolution of the magnetic field is described by 
equation (\ref{magfluxeq}) with the boundary conditions
$\psi(r_o)=B_o r_o e^{i\phi}$ and $d\psi(r_o)/dr=B_o e^{i\phi}$ at
$r=r_o$.
In a steady state, we can set $\psi=F(r)e^{in\phi}$ since the coefficients
of the equation do not depend explicitly on $\phi$.
Note that with the given initial magnetic field and boundary conditions,
only the solution with $n=1$ is allowed.
Then, the equation for $F$ is given as
\begin{equation}
{dF\over dr^2}+{1\over r}{dF\over dr}-{1\over r^2}F-i{Q\over r_o^2}F=0,
\label{magffuneq}
\end{equation}
where
\begin{equation}
Q = \cases{q, &for $0<r<r_o/2$\cr
q(r_o/r-1), & for $r_o/2<r<r_o$\cr
0, & for $r>r_o$\cr}
\label{qeq}
\end{equation}
with $q=2Ar^2_o/\eta$.
Here, $q$ is the magnetic Reynolds number within a factor of two.

We solve equation (\ref{magffuneq}) for  $F$ numerically.
In Figure 5, we plot the resulting magnetic field lines for the case with
$q=10$ (highly diffusive case) and for the case with $q=10^3$ (quasi-adiabatic
case).
From dissipation tests we estimate the magnetic Reynolds number of the vortex 
in our time-dependent simulations to be larger
than $10^3$ (see \cite{ryujf95}; Paper I). 
Indeed the fields in Figure 4b resemble those in a vortex simulation
with magnetic Reynolds number $10^3$ reported in \cite{weiss66}.
The steady-state solution we just described allows only for diffusion
of magnetic flux, since no change in the field topology is permitted.
It allows us to see that such a steady-state, diffusive description
does not account for flux expulsion from the vortex.
If that were the case
the magnetic field lines
in Figure 4b should have a structure like those in the bottom of Figure 5.
Actually, their topologies are fundamentally different in the sense 
that field in Figure 5 threads completely through the vortex, while
it does not in Figure 4b (or any similar figure showing additional field lines for this
simulation).
The reason is that the field in the time dependent simulation
is subjected to reconnective instabilities that isolate and
then destroy magnetic flux in the interior of the vortex.

\clearpage

\begin{deluxetable}{ccccccccc}
\footnotesize
\tablecaption{Summary of MHD KH Simulations \label{tbl-1}}
\tablehead{
\colhead{Case \tablenotemark{a}} & 
\colhead{$B_o = c_a/c_s$\tablenotemark{b}} &
\colhead{$\theta$} & 
\colhead{$B_{po} = B_o\cos{\theta}$} &
\colhead{$M_A$\tablenotemark{b}} & 
\colhead{$\beta_0$\tablenotemark{b}}  &
\colhead{$t_g$\tablenotemark{c}} &
\colhead{End Time\tablenotemark{c}} &
\colhead{$N_x$\tablenotemark{d}}
} 
\startdata
1\tablenotemark{e} &  0.4 &0\arcdeg & 0.4 &2.5&7.5&3.79
& 20$\tau$ & 512\tablenotemark{e}  \nl
2\tablenotemark{e} &  0.2  & 0\arcdeg & 0.2&5.0 & 30&1.86
& 20$\tau$ & 512\tablenotemark{e} \nl
3 & 0.2  & 45\arcdeg & 0.14&5.0 & 30&1.71 & 30$\tau$ & 512\nl
4 &  0.14 & 0\arcdeg & 0.14&7.07 & 60&1.71 & 30$\tau$ & 512\nl
5 &   0.2 & 90\arcdeg & 0.0&5.0 & 30&1.59 & 20$\tau$ & 256\nl
6h &  0.2 &85\arcdeg  &0.02&5.0 & 30&1.59 & 50$\tau$ & 512 \nl
6m &  0.2 &85\arcdeg  &0.02&5.0 & 30&1.59 & 30$\tau$ & 256 \nl
7 &  0.02 &0\arcdeg & 0.02&50   & 3000&1.59 & 30$\tau$ & 256\nl
8 &  0.04 &0\arcdeg & 0.04&25   & 750&1.67 & 30$\tau$ & 256\nl
9 &  0.07 &0\arcdeg & 0.07&14.3 & 245&1.67 & 20$\tau$ & 256\nl
 
\enddata

\tablenotetext{a}{All models have used $\gamma$ = 5/3,
$M = U_0/c_s = 1$, $c_s = 1$, $L = 2.51$, $a = L/25$.}
\tablenotetext{b}{The Alfv\'en speed, $c_a = B/\sqrt{\rho}$,
Alfv\'en Mach number, and
$\beta_0 = p_g/p_b = (2/\gamma) (M_A/M)^2$ here refer to the
total initial magnetic field strength, not just that projected
onto the plane of the flow.}
\tablenotetext{c}{The growth time is an approximation
to the inverse linear growth rate; namely $t_g = 1/\Gamma$; $\tau = t/t_g$.} 
\tablenotetext{d}{Computations were carried out on a square grid of
the size indicated, with $N_y = N_x$. Periodic boundaries were
assumed for $x$ and
reflecting boundaries were assumed for $y$.}
\tablenotetext{e}{Cases 1 and 2 were presented and discussed in Paper I.
They are cited here for reference. Each was computed with 2 or more
numerical resolutions, with the largest listed here.}
 
\end{deluxetable}

\clearpage

\begin{center}
{\bf FIGURE CAPTIONS}
\end{center}
\begin{description}

\item[Fig.~1]
{Cartoon illustrating the computational setup for these simulations.
There is a central shear layer separating uniform flows. Except for the
initial perturbation the flow conditions are otherwise uniform. The
magnetic field projects onto the computational plane at an angle, $\theta$.}

\item[Fig.~2]
{Time evolution of the high resolution simulations (Cases 3, 4, 6h), plus
the medium resolution simulation Case 6m. Shown are the normalized
total thermal, kinetic and magnetic energies, as well as the minimum
value of the plasma $\beta$ parameter at each time.
To emphasize the minimal importance of the transverse magnetic
field component, $B_z$, ``reduced'' energies are shown that
exclude $B_z$. }

\item[Fig.~3]
{Same as Figure 2, except for medium resolution simulations
(Cases 5, 6m 7, 8, 9).}

\item[Fig.~4a]
{Greyscale images of the Case 6h vorticity component
normal to the computational
plane at $\tau = 5$ (top) and $\tau = 30$ (bottom).
To facilitate visualization
of structures, the periodic space has been repeated once. The vorticity is
everywhere positive. High values take high tones.}

\item[Fig.~4b]
{Case 6h magnetic field lines projected onto the computational plane
plane at $\tau = 5$ (top) and $\tau = 30$ (bottom). }

\item[Fig.~5]
{Magnetic field lines from an analytical model to study diffusive
flux expulsion from a steady eddy with $q=10$ (top) and $q=1000$
(bottom). See the text for the definition of $q$.}

\item[Fig.~6]
{Grayscale snapshots of the magnetic pressure distributions for Case 3 
at $\tau = 5,~10$ and 30, showing the stages in disruption of the
cat's eye vortex by the magnetic field. High values take high tones.}

\item[Fig.~7]
{Inverted grayscale image of $\log{\beta} = p/p_b$ for Case 3 at
$\tau = 5$. Projected magnetic field lines are overlaid. The
minimum $\beta \sim 0.55$ in the strong flux tube between
vortices, while the maximum $\beta \sim 10^6$
near the center of the cat's eye.}

\item[Fig.~8]
{The magnetic field structure for Case 3 at $\tau = 7.5$, just as
substantial magnetic reconnection is underway. Projected field
lines are shown, along with arrows that represent magnetic
tension forces. The maximum magnetic tension force is 1.74}

\item[Fig.~9]
{Inverted grayscale image of the gas entropy (top) and
electrical current (bottom)
distributions for the
Case 3 at $\tau = 7.5$. Magnetic field lines are also laid on top of
the current distribution to emphasize the relationships. High values take
low tones.}

\end{description}


\clearpage

\begin{thebibliography}{}


\bibitem[Balbus \& Hawley (1991)] {balhal91}
Balbus, S. \& Hawley, J. 1991, \apj, 376, 214

\bibitem[Bayly (1986)]{bayly86}
Bayly, B.J. 1986, Phys. Rev. Lett., 57, 2160

\bibitem[Bayly, Orszag \& Herbert (1988)]{boh88}
Bayly, B.J., Orszag, S.A., \& Herbert, T. 1988, Ann. Rev. Fluid Mech.,
20, 359



\bibitem[Biskamp \& Welter (1989)] {biswel89} 
Biskamp, D. \& Welter, H. 1989, Phys. Fluids B 1, 1964

\bibitem[Biskamp (1993)] {bis93}
Biskamp, D. 1993, ``Nonlinear Magnetohydrodynamics'', (Cambridge University
Press: Cambridge)

\bibitem[Biskamp (1994)] {bis94}
Biskamp, D. 1994, Physics Rept., 237, 179





\bibitem[Cattaneo \& Vainstein (1991)] {catvain91} 
Cattaneo, F. \& Vainstein, S. I. 1991, \apj, 376, L21


\bibitem[Chandrasekhar (1961)] {chan61} 
Chandrasekhar, S. 1961, ``Hydrodynamic \& Hydromagnetic Stability'', (Oxford
University Press: New York)

\bibitem[Corcos \& Sherman (1984)] {corsher84} 
Corcos, G. M. \& Sherman, F. S., 1984, J. Fluid Mech., 139, 29


\bibitem[Frank \etal (1996)] {fran96}
Frank, A., Jones, T. W., Ryu, D., \& Gaalaas, J. B. 1996, \apj, 460, 777
(Paper I)



\bibitem[Harten (1983)] {harten83}
Harten, A. 1983, J.  Comp. Phys., 49, 357

\bibitem[Jun, Norman \& Stone (1995)] {junetal95}
Jun, B. I., Norman, M. L., \& Stone, J. M. 1995, \apj, 453, 332



\bibitem[Landau, Lifshitz, \& Pitaevskii (1984)] {landau84}
Landau, L. D., Lifshitz, E. M., \& Pitaevskii, L. P. 1984,
``Electrodynamics of Continuous Medium'', 
(Pergamon Press: Oxford)


\bibitem[Malagoli, Bodo \& Rosner (1996)] {maletal96} 
Malagoli, A., Bodo, G., \& Rosner R. 1996 \apj, 456, 708 (MBR)




\bibitem[Miura (1984)] {miura84} 
Miura, A. 1984, JGR, 89, 801


\bibitem[Miura (1990)] {miura90} 
Miura, A. 1990, Geophys. Res. Lett., 17, 749


\bibitem[Miura \& Pritchett (1982)] {miupr82} 
Miura, A. \& Pritchett, P. L. 1982, JGR, 87, 7431 (MP)

\bibitem[Moffatt (1978)] {moff78} 
Moffatt, H. K. 1978, ``Magnetic Field Generation in Electrically Conducting
Fluids'', (Cambridge University Press: Cambridge)

\bibitem[Nordlund \etal (1992)] {noret92} 
Nordlund, \AA, \etal 1992, \apj, 392, 647

\bibitem[Parker (1994)] {park94}
Parker, E. N. 1994, ``Spontaneous Current Sheets in Magnetic Fields'',
(Oxford University Press: New York)

\bibitem[Pedelty \& Woodward (1991)] {pedw91}
Pedelty, J. A. \& Woodward, P. R. 1991, J. Fluid Mech., 225, 101

\bibitem[Pierrehumbert (1986)]{pierr86}
Pierrehumbert, S. 1986, Phys. Rev. Lett., 57, 2157

\bibitem[Porter \& Woodward (1994)] {porwood94}
Porter, D. H. \& Woodward, P R. 1994 \apjs, 93, 309



\bibitem[Priest (1984)] {priest84} 
Priest, E. R. 1984, ``Solar Magnetohydrodynamics'', (D. Reidel Publ. Co.:
Dortrecht)

\bibitem[Ryu \& Jones  (1995)] {ryuj95} 
Ryu, D. \&  Jones, T. W. 1995, \apj, 442, 228

\bibitem[Ryu, Jones \& Frank (1995)] {ryujf95} 
Ryu, D., Jones, T. W., \& Frank, A. 1995, \apj, 452, 785




\bibitem[Weiss (1966)] {weiss66}
Weiss, N. O. 1966, Proc. Roy. Soc. London, A, 293, 310



\bibitem[Zimmer, Lesch \& Birk (1997)] {zim96}
Zimmer, F., Lesch, H. \& Birk, G. T. 1997, A\& A, (in press)

\end{thebibliography}
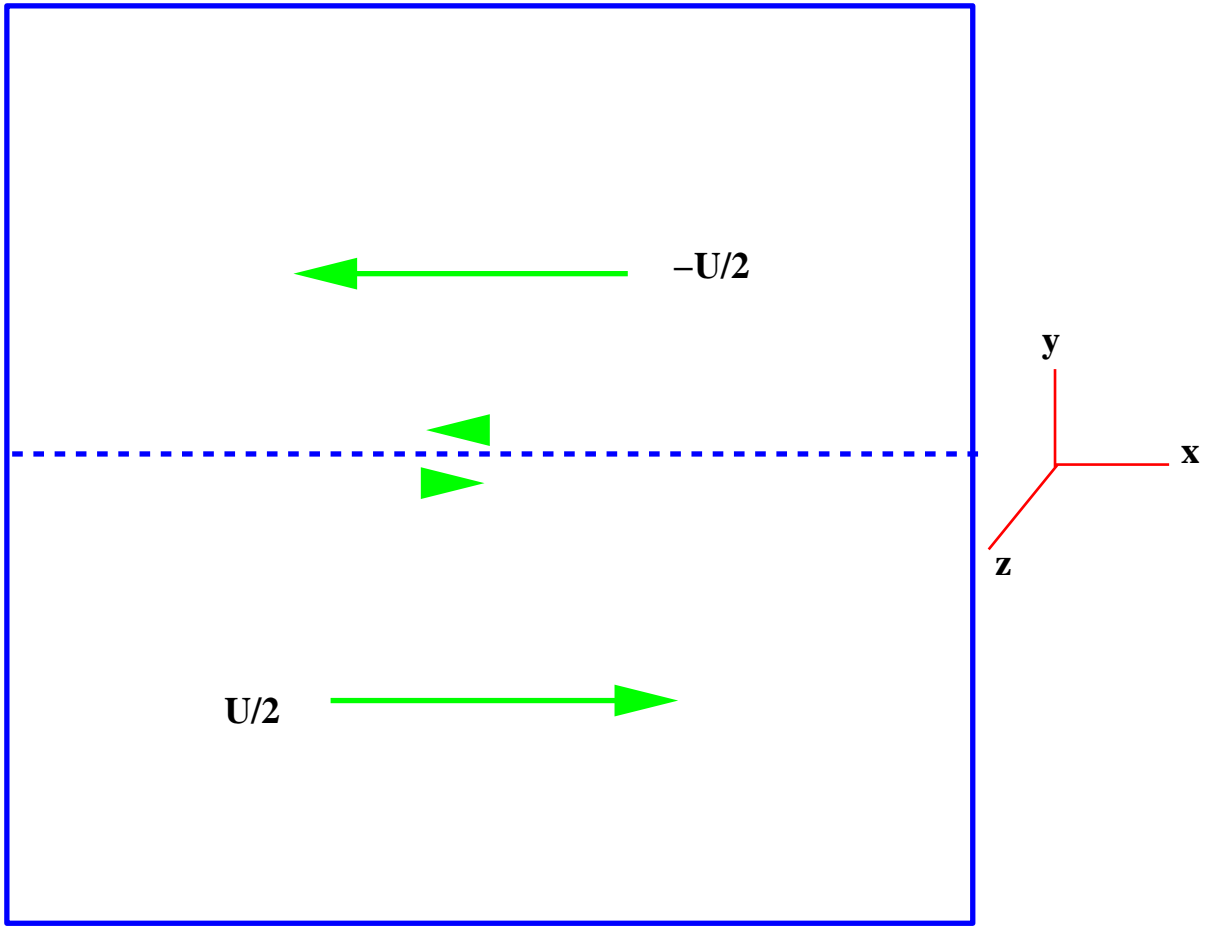
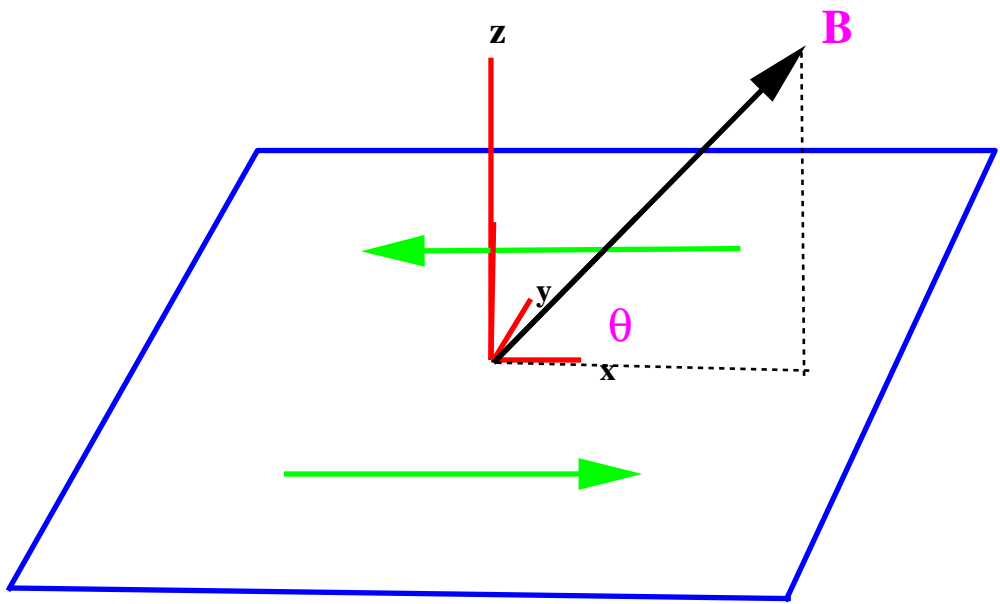
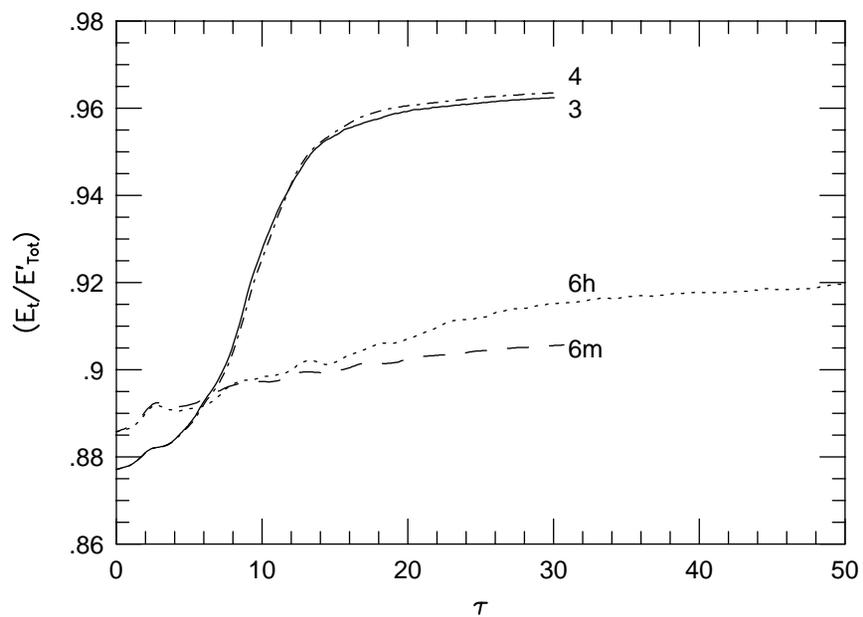
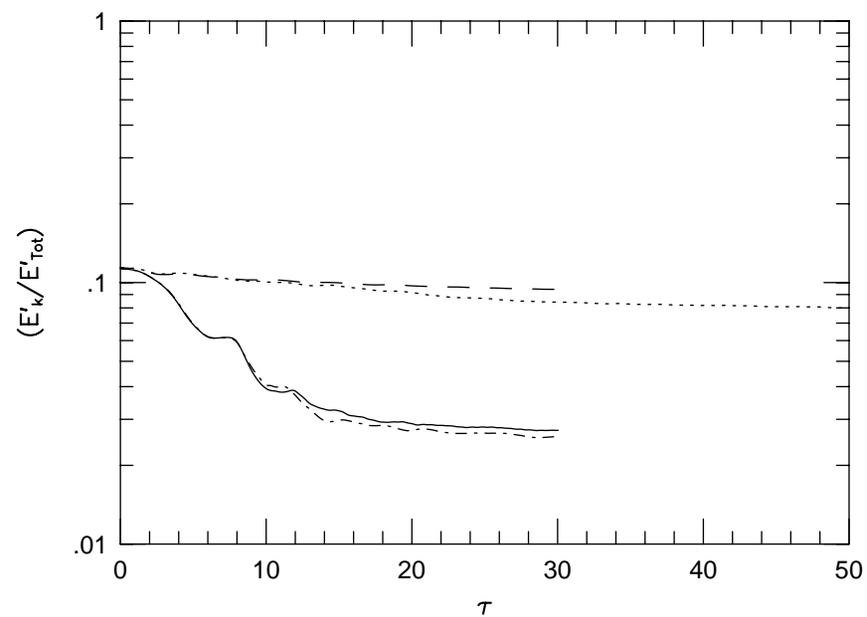
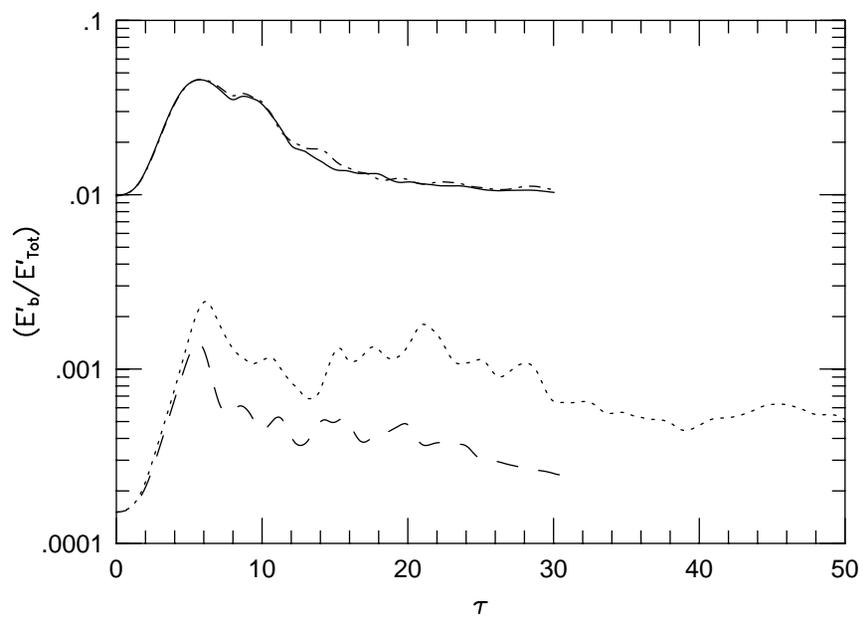
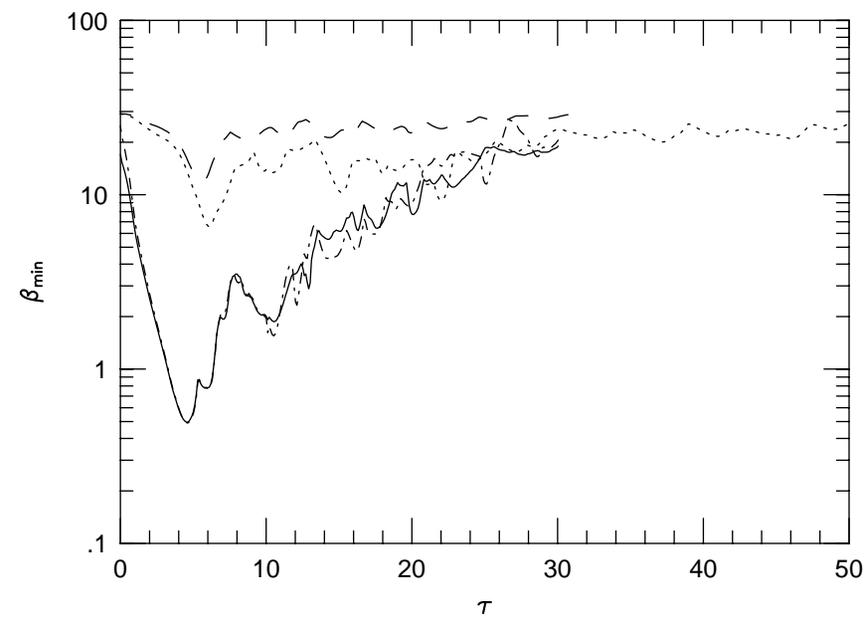
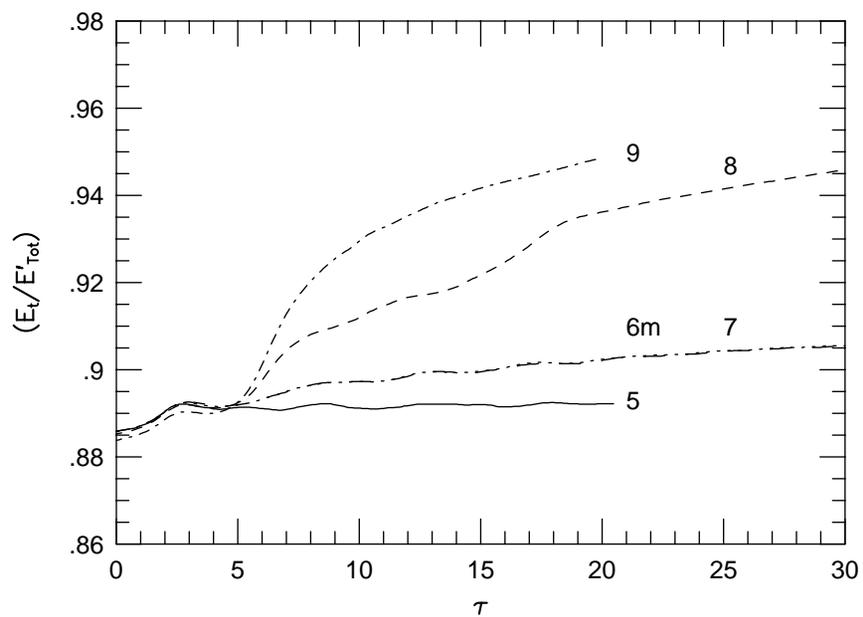
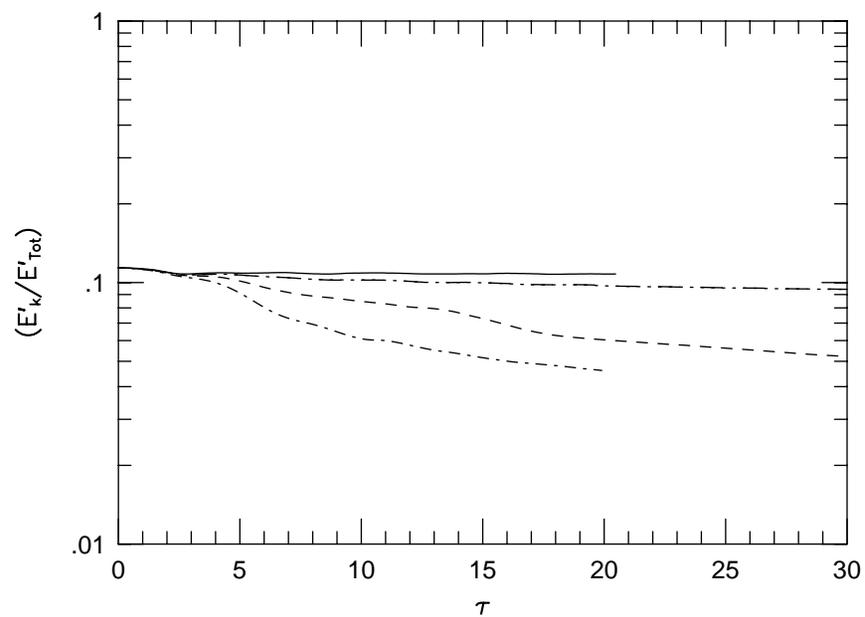
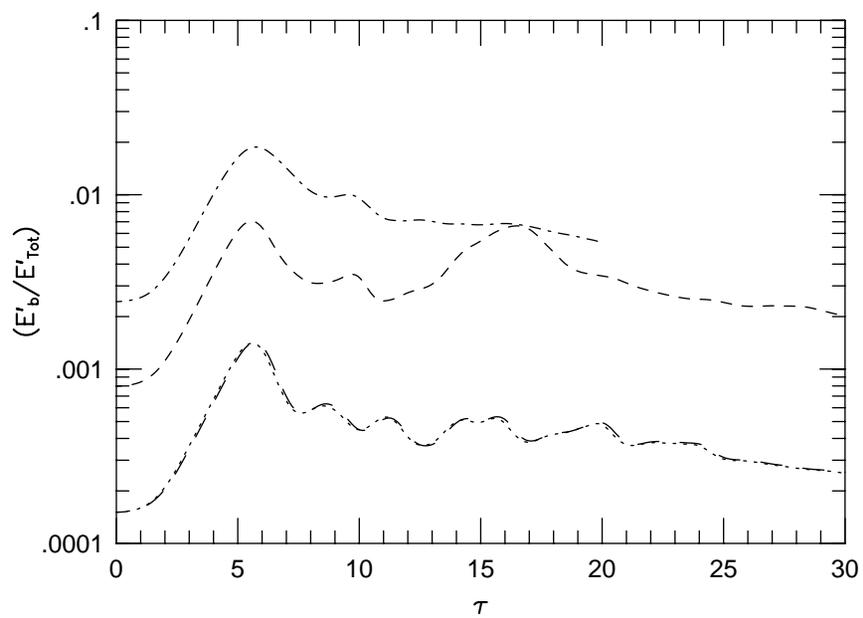
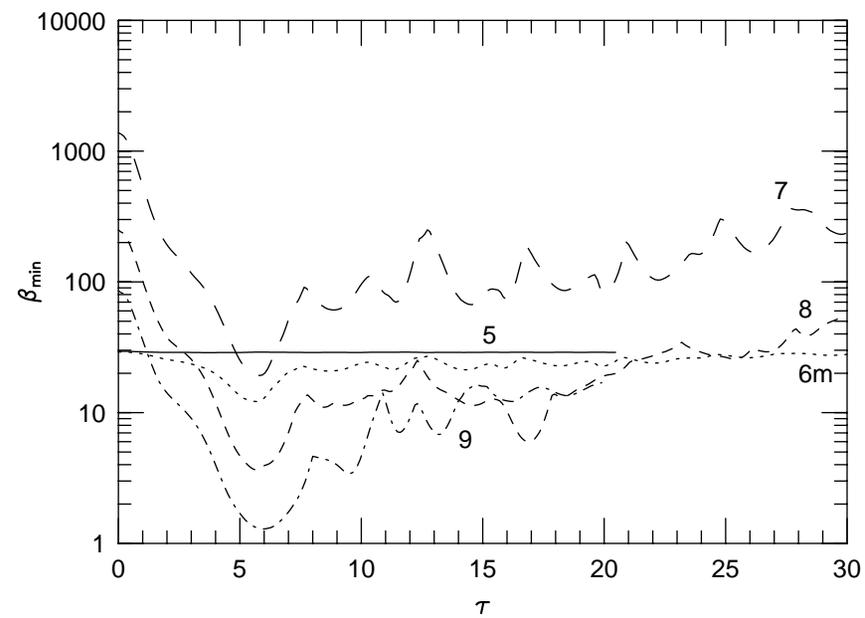
\end{document}